\documentclass[12pt]{article}
\newcommand{\beq}{\begin{equation}}
\newcommand{\eeq}{\end{equation}}
\newcommand{\beqa}{\begin{eqnarray}}
\newcommand{\eeqa}{\end{eqnarray}}
\newcommand{\beqar}{\begin{eqnarray*}}
\newcommand{\eeqar}{\end{eqnarray*}}

\newcommand{\labell}[1]{\label{#1}} 
\newcommand{\reef}[1]{(\ref{#1})}

\newcommand{\ssc}{\scriptscriptstyle}
\newcommand{\eg}{{\it e.g.,}\ }
\newcommand{\ie}{{\it i.e.,}\ }

\newcommand{\norm}[1]{\raise.3ex\hbox{:}#1\raise.3ex\hbox{:}}

\newcommand{\Tr}{{\rm Tr}}
\newcommand{\STr}{{\rm STr}}

\newcommand\hi{{\rm i}}

\newcommand\prt{\partial}

\newcommand{\al}{\alpha}

\newcommand{\veps}{\varepsilon}
\newcommand\vareps{\varepsilon}

\newcommand{\eps}{\epsilon}

\renewcommand{\l}{\lambda}

\newcommand{\s}{\sigma}

\newcommand\cL{{\cal L}}
\newcommand\cH{{\cal H}}

\newcommand{\phd}{\dot{\phi}}
\newcommand{\tOmega}{\widetilde{\Omega}}

\newcommand\ls{\ell_s}

\newcommand\signot{\sigma_\infty}

\renewcommand{\S}{{\rm S}}
\newcommand{\AdS}{{\rm AdS}}
\newcommand\Pp{P_\phi}

\def\abstracts#1#2#3{{
	\centering{\begin{minipage}{4.5in}\baselineskip=10pt\footnotesize
	\parindent=0pt #1\par 
	\parindent=15pt #2\par
	\parindent=15pt #3
	\end{minipage}}\par}} 
\newcommand{\textlineskip}{\baselineskip=13pt} 

\renewcommand{\l}{\lambda}
\newcommand\sig{\sigma}
\newcommand\hx{{\hat x}}
\newcommand{\bear}{\begin{eqnarray}}
\newcommand{\eear}{\end{eqnarray}}
\newcommand{\NN}{{\rm N}} 
\newcommand\mathC{{\bf C}} 
\newcommand\mathR{I\!\!R}

\begin{document}
\normalsize\textlineskip


\thispagestyle{empty}
\setcounter{page}{1}

\leftline{\hfill\small hep--th/0106178}\nopagebreak
\vskip -.6ex

\vspace*{0.48truein}

\centerline{\bf NONABELIAN D-BRANES AND NONCOMMUTATIVE GEOMETRY}
\vspace*{0.4truein}
\centerline{\footnotesize ROBERT MYERS}
\vspace*{0.02truein}
\centerline{\footnotesize\it Department of Physics, McGill University,
3600 University Street}
\baselineskip=10pt
\centerline{\footnotesize\it Montr\'eal, Qu\'ebec, H3A 2T8, Canada}

\vspace*{0.22truein}
\abstracts{We discuss the nonabelian world-volume action which governs the
dynamics
of N coincident D$p$-branes. In this theory, the branes' transverse
displacements are described by matrix-valued scalar fields, and so this is
a natural physical framework for the appearance of noncommutative geometry.
One example is the
dielectric effect by which D$p$-branes may be polarized into
a noncommutative geometry by external fields.
Another example is the appearance of noncommutative geometries in the
description of intersecting D-branes of differing dimensions, such as D-strings
ending on a D3- or D5-brane. We also describe the related physics
of giant gravitons.}{}{}


\vspace*{18pt}
The idea that noncommutative geometry should play a role in
physical theories is an old one \cite{old,old2}. Suggestions have been
made that such noncommutative structure may resolve the
ultraviolet divergences of quantum field theories, or appear
in the description of spacetime geometry at the Planck scale.
In the past few years, it has also become a
topic of increasing interest to string theorists. From one point of
view, the essential step in realizing a noncommutative geometry is
replacing the spacetime coordinates by noncommuting operators:
$x^\mu\rightarrow\hx^\mu$. In this replacement, however, there remains
a great deal of freedom in defining the nontrivial commutation relations
which the operators $\hx^\mu$ must satisfy. Some explicit choices that have
appeared in physical problems are as follows:
\vskip 1ex
\noindent\textit{(i) Canonical commutation relations}:
\[
[\hx^\mu,\hx^\nu]=i\theta^{\mu\nu}\qquad \theta^{\mu\nu}\in \mathC
\]
\noindent Such algebras have appeared in the Matrix theory description of
planar D-branes \cite{matrix} --- for a review, see \cite{watirev}.
This work also stimulated an ongoing investigation
by string theorists of noncommutative field theories which arise in the
low energy limit of a planar D-brane
with a constant B-field flux --- see, \eg \cite{cds,dh,seiwit}.
\vskip 1ex
\noindent\textit{(ii) Quantum space relations}:
\[
\hx^\mu\,\hx^\nu= q^{-1}\, R^{\mu\nu}{}_{\rho\tau}\,\hx^\rho\,\hx^\tau
\qquad R^{\mu\nu}{}_{\rho\tau}\in \mathC
\]
\noindent These algebras received some attention from physicists in the early
1990's --- see, \eg \cite{zum1,zum2} --- and have appeared more recently in the geometry
of the moduli space of $N$=4 super-Yang-Mills theory \cite{leigh}.
\vskip 1ex
\noindent\textit{(iii) Lie algebra relations}:
\[
[\hx^\mu,\hx^\nu]=if^{\mu\nu}{}_\rho\,\hx^\rho\qquad f^{\mu\nu}{}_\rho\in
\mathC
\]
\noindent Such algebras naturally arise in the description of fuzzy spheres
as was discovered in early attempts to quantize
the supermembrane \cite{hoppe,hoppen}. These noncommutative geometries have
also been
applied in Matrix theory to describe spherical D-branes \cite{wati1,wati2}.
In string theory, these noncommutative descriptions of spheres also arise in
various contexts in the physics of D-branes, as will be discussed below.
\vskip 1ex

For a system of N (nearly) coincident D-branes, the
transverse displacements are described by a set of scalar fields,
which are matrix-valued in the adjoint representation of U(N).
Hence, noncommutative geometries with a Lie-algebra structure
appear very naturally in the physics of D-branes.
The appearance of a nonabelian U(N) gauge
symmetry in the world-volume theory of N coincident D-branes \cite{wite} is, of course,
one of the most remarkable aspects of the D-brane story \cite{dbrane1,dbrane2}.
It lies at the heart of such recent developments as the entropy counting of
near-BPS black holes \cite{peet} and the AdS/CFT correspondence \cite{revue}.
Progress has recently been made on constructing the world-volume action that
controls the dynamics of this nonabelian theory \cite{die,wati3}.
In particular, one finds that this action includes a wide variety of
new nonderivative terms for the world-volume scalars. Amongst these
interactions are couplings by which the nonabelian D-branes can interact with
all of the Ramond-Ramond potentials of any form degree. Further,
there is an interesting ``dielectric effect'' \cite{die} in which
the D-branes are polarized into a higher dimensional noncommutative
geometry by nontrivial background fields.

An outline of this paper is as follows: We begin in section 1 
with a discussion of the nonabelian D-brane action. Section 2 
presents an outline of the dielectric effect for D-branes. Section 3
describes the related physical effect by which branes carrying momentum
expand in AdS$_m\times\S^n$ backgrounds, producing giant gravitons.
Finally, section 4 
gives a discussion of how noncommutative
geometries can arise in the description of intersecting branes.
Sections 1 and 2 
are essentially a summary of the
material appearing in ref.~\cite{die}. Section 3 describes that
in ref.~\cite{goliath} and  section 4 
describes that for refs.~\cite{bion} and \cite{fiv}. We direct the
interested reader to these papers for a more detailed presentation
of the associated works.

\vskip 3ex
\centerline{\bf 1. Nonabelian D-brane action}   

Within the framework of perturbative string theory,
a D$p$-brane is a
($p+1$)-dimensional extended surface in spacetime which supports
the endpoints of open strings \cite{dbrane1,dbrane2}.
The massless modes of this open string theory form a supersymmetric
U(1) gauge theory with a vector $A_a$, $9-p$ real scalars $\Phi^i$ and
their superpartner fermions --- for the most part,
the latter are ignored throughout the
following discussion. 
At leading order, the low-energy action corresponds to the
 dimensional reduction of that for ten-dimensional
U(1) super-Yang-Mills theory. However, as usual in string theory, there are
higher order $\alpha'=\ls^2$ corrections --- $\ls$ is the string length
scale. For constant field strengths, these stringy corrections can be resummed
to all orders, and the resulting action takes the Born-Infeld form \cite{bin}
\begin{equation}
\labell{biact}
S_{BI}=-T_p \int d^{p+1}\sig\ \left(e^{-\phi}\sqrt{-det(P[G+B]_{ab}+
\l\,F_{ab})}\right)
\end{equation}
where $T_p$ is the D$p$-brane tension 
and $\l$ denotes the inverse of the (fundamental) string tension, \ie
$\l=2\pi\ls^2$. This Born-Infeld action describes
the couplings of the D$p$-brane to the
massless Neveu-Schwarz fields of the bulk closed string theory,
\ie the metric, dilaton and Kalb-Ramond two-form.
The interactions with the massless Ramond-Ramond (RR) fields are incorporated
in a second part of the action, the Wess-Zumino term\cite{mike,cs,cs2}
\begin{equation}
S_{WZ}=\mu_p\int P\left[ \sum C^{(n)}\,e^B\right]e^{\l\,F}
\labell{csact}
\end{equation}
where $C^{(n)}$ denote the $n$-form RR potentials.
Eq.~\reef{csact} shows that a D$p$-brane is naturally
charged under the ($p$+1)-form RR potential with charge $\mu_p$,
and supersymmetry dictates that $\mu_p=\pm T_p$.
If the D$p$-brane carries a flux of $B+F$, it will also act as a charge source
for RR potentials with a lower form degree \cite{mike}.
Such configurations represent bound states
of D-branes of different dimensions\cite{wite}.

In both of the expressions above, the symbol $P[\ldots]$
denotes the pull-back of the bulk spacetime tensors to the D-brane world-volume.
Thus the Born-Infeld action \reef{biact} has a geometric interpretation,
\ie it is
essentially the proper volume swept out by the D$p$-brane, which is indicative of
the fact that D-branes are actually dynamical objects. This dynamics becomes
more evident with an explanation of the static gauge choice implicit in
constructing the above action. To begin, we employ spacetime diffeomorphisms to position
the world-volume on a fiducial surface defined as $x^i=0$ with $i=p+1,\ldots,9$.
With world-volume diffeomorphisms, we then match the world-volume coordinates with
the remaining spacetime coordinates on this surface, $\sig^a=x^a$ with $a=0,1,\ldots,p$.
Now the world-volume scalars $\Phi^i$ play the role of describing the transverse
displacements of the D-brane, through the identification
\begin{equation}
\labell{match}
x^i(\sig)=2\pi\ls^2\Phi^i(\sig)\qquad {\rm with}\ i=p+1,\ldots,9.
\end{equation}
With this identification the general formula for the pull-back reduces to
\bear
\labell{pull}
P[E]_{ab}&=&E_{\mu\nu}{\prt x^\mu\over\prt\sig^a}{\prt x^\nu\over\prt\sig^b}\\
&=&E_{ab}+\l\,E_{ai}\,\prt_{b}\Phi^i+
\l\,E_{ib}\, \prt_a\Phi^i+
\l^2\,E_{ij}\prt_a\Phi^i\prt_b\Phi^j\ .
\nonumber
\eear
In this way, the expected kinetic terms for the scalars emerge to
leading order in an expansion of the Born-Infeld action \reef{biact}.
Note that our conventions are such that both the gauge fields
and world-volume scalars have the dimensions of $length^{-1}$
--- hence the appearance of the string scale in eq.~\reef{match}.

As N parallel D-branes approach each other, the ground state modes of
strings stretching between the different D-branes become massless. These
extra massless states carry the appropriate charges to fill out
representations under a U(N) symmetry. Hence the U(1)$^{\rm N}$ of the individual D-branes
is enhanced to the nonabelian group U(N) for the coincident D-branes \cite{wite}.
The vector $A_a$ becomes a nonabelian gauge field
\begin{equation}
A_a=A_a^{(n)}T_n,
\qquad F_{ab}=\prt_a A_b-\prt_b A_a+i[A_a,A_b]
\labell{gauge}
\end{equation}
where $T_{n}$ are $\NN^2$ hermitian generators
with $\Tr(T_{n}\,T_m)=\NN\,\delta_{n m}$.
The scalars $\Phi^i$ also transform in the adjoint of U(N)
with covariant derivatives
\begin{equation}
D_a\Phi^i=\prt_a\Phi^i+i[A_a,\Phi^i]\ .
\labell{scalder}
\end{equation}

Understanding how to accommodate this U(N) gauge symmetry in the world-volume
action is an interesting puzzle. For example, the geometric meaning or even the
validity of eq.~\reef{match} seems uncertain when the scalars on the right hand
side are matrix-valued. In fact, this identification does remain roughly correct.
Some intuition comes from the case where the scalars are commuting matrices and
the gauge symmetry can be used to simultaneously diagonalize all of
them. In this case,
one interpretes the N eigenvalues of the diagonal $\Phi^i$ as representing the
displacements of the N constituent D-branes --- see, \eg \cite{watirev}.
Of course, to describe noncommutative
geometries, we will be more interested in the case where the scalars do not
commute and so cannot be simultaneously diagonlized.

Refs.~\cite{die} and \cite{wati3} recently made progress in constructing the
world-volume
action describing the dynamics of nonabelian D-branes. The essential strategy
in both of these papers was to
construct an action which was consistent with the familiar string theory
symmetry of T-duality \cite{tdual}. Acting on D-branes, T-duality acts to
change the dimension
of the world-volume \cite{dbrane1,dbrane2}. The two possibilities are:
{\it (i)} if a coordinate transverse to the D$p$-brane, \eg $y=x^{p+1}$, is
T-dualized, it becomes a D$(p+1)$-brane where $y$ is now the
extra world-volume direction; and {\it (ii)} if a world-volume coordinate
on the D$p$-brane, \eg $y=x^{p}$, is
T-dualized, it becomes a D$(p-1)$-brane where $y$ is now an
extra transverse direction. Under these transformations, the role of the
corresponding world-volume fields change as
\begin{equation}
(i)\ \Phi^{p+1}\,\rightarrow\, A_{p+1}\ ,
\qquad\qquad
(ii)\ A_p\,\rightarrow\,\Phi^p\ ,
\labell{rule1}
\end{equation}
while the remaining components of $A$ and scalars $\Phi$ are left
unchanged. Hence in constructing the nonabelian action, one can begin
with the D9-brane theory, which contains no scalars since the world-volume
fills the entire spacetime. In this case, the nonabelian extension of
eqs.~\reef{biact} and \reef{csact} is given by simply introducing an overall
trace over gauge indices of the nonabelian field strengths appearing in the
action \cite{moremike}. Then applying T-duality transformations on $9-p$
directions yields
the nonabelian action for a D$p$-brane. Of course, in this construction, one
also T-dualizes the background supergravity fields according to the known
transformation rules \cite{tdual,Tdual,Tdual1,Tdual2}.
As in the abelian theory, the result for nonabelian action
has two distinct pieces \cite{die,wati3}: the Born-Infeld term
\bear
{S}_{BI}&=&-T_p \int d^{p+1}\sigma\,\STr\left(e^{-\phi}\, \sqrt{\det(Q^i{}_j)}
\right.\nonumber\\
&&\left.\qquad\qquad
\times\ \sqrt{-\det\left(
P\left[E_{ab}+E_{ai}(Q^{-1}-\delta)^{ij}E_{jb}\right]+
\l\,F_{ab}\right)}
\right),
\labell{finalbi}
\eear
with
$E_{\mu\nu}=G_{\mu\nu}+B_{\mu\nu}$
and $Q^i{}_j\equiv\delta^i{}_j+i\lambda\,[\Phi^i,\Phi^k]\,E_{kj}$;
and the Wess-Zumino term
\begin{equation}
S_{WZ}=\mu_p\int \STr\left(P\left[e^{i\l\,\hi_\Phi \hi_\Phi} (
\sum C^{(n)}\,e^B)\right]
e^{\l\,F}\right)\ .
\labell{finalcs}
\end{equation}

Let us enumerate the nonabelian features of this action:
\vskip 1ex
\noindent 1.~\textit{Nonabelian field strength}:
The $F_{ab}$ appearing explicitly in both terms
is now nonabelian, of course.
\vskip 1ex
\noindent 2.~\textit{Nonabelian Taylor expansion}:
The bulk supergravity fields
are in general functions of all of the spacetime coordinates, and so
in the action (\ref{finalbi},\ref{finalcs}), they are implicitly functionals
of the nonabelian scalars.
For example, the metric functional appearing in
the D-brane action would be given by a {nonabelian} Taylor expansion
\bear
G_{\mu\nu}&=&\exp\left[\l\Phi^i\,{\prt_{x^i}}\right]G^0_{\mu\nu}
(\sigma^a,x^i)|_{x^i=0}
\labell{slick}\\
&=&\sum_{n=0}^\infty {\l^n\over n!}\,\Phi^{i_1}\cdots\Phi^{i_n}\,
(\prt_{x^{i_1}}\cdots\prt_{x^{i_n}})G^0_{\mu\nu}
(\sigma^a,x^i)|_{x^i=0}\ .
\nonumber
\eear
\vskip 1ex
\noindent 3.~\textit{Nonabelian Pullback}:
As was noted in refs.~\cite{hull,dorn}, the pullback of various spacetime tensors
to the world-volume must now involve covariant derivatives of the nonabelian scalars
in order to be consistent with the U(N) gauge symmetry. Hence eq.~\reef{pull} is
replaced by
\begin{equation}
P[E]_{ab}=E_{ab}+\l\,E_{ai}\,D_{b}\Phi^i+
\l\,E_{ib}\, D_a\Phi^i+
\l^2\,E_{ij}D_a\Phi^iD_b\Phi^j\ .
\labell{pulla}
\end{equation}
\noindent 4.~\textit{Nonabelian Interior Product}:
In the Wess-Zumino term \reef{finalcs},
$\hi_\Phi$ denotes the interior product with $\Phi^i$ regarded
as a vector in the transverse space, {\it e.g.}, acting on an $n$-form
$C^{(n)}={1\over n!}C^{(n)}_{\mu_1\cdots\mu_n} dx^{\mu_1}\cdots dx^{\mu_n}$,
we have
\begin{equation}
\hi_\Phi\hi_\Phi C^{(n)}={1\over2(n-2)!}\,[\Phi^i,\Phi^j]\,
C^{(n)}_{ji\mu_3\cdots\mu_n}dx^{\mu_3}\cdots dx^{\mu_n}
\ . \labell{inside}
\end{equation}
Note that acting on forms, the interior product is an anticommuting operator
and hence for an ordinary vector (\ie a vector $v^i$ with values in
$\mathR^{9-p}$):
$\hi_v\hi_vC^{(n)}=0$. It is only because the scalars $\Phi$ are matrix-valued
that eq.~\reef{inside} yields a nontrivial result.
\vskip 1ex
\noindent 5.~\textit{Nonabelian Gauge Trace}:
As is evident above, both parts of the action
are highly nonlinear functionals of the nonabelian fields, and so
eqs.~\reef{finalbi}
and \reef{finalcs} would be incomplete without a precise definition for the ordering of
these fields under the gauge trace. Above, $\STr$ denotes a maximally symmetric trace.
To be precise, the trace includes a symmetric average over all orderings of $F_{ab}$,
$D_a\Phi^i$, $[\Phi^i,\Phi^j]$ and the individual $\Phi^k$ appearing in the
nonabelian Taylor expansions of the background fields. This choice matches
that inferred from Matrix theory \cite{wati4}, and a similar symmetrization
arises in the leading order analysis of the boundary $\beta$ functions
\cite{dorn}. Finally we should note that
with this definition an expansion of the Born-Infeld term \reef{biact} does agree
with the string theory to fourth order in $F$ \cite{arkady,ark2}, however, it does not
seem to capture the full physics of the nonabelian fields in the infrared limit
\cite{wati5}. Rather at sixth order, additional terms involving commutators of
field strengths must be added to the action \cite{bain}.
\vskip 1ex

Some other general comments on the nonabelian action are as follows:
In the Born-Infeld term \reef{finalbi}, there are now two determinant factors
as compared to one in the abelian action \reef{biact}. The second determinant
in eq.~\reef{finalbi} is a slightly modified version of that in eq.~\reef{biact}.
One might think of this as the kinetic factor, since to leading order in the low
energy expansion, it yields the familiar kinetic terms for the gauge field
and scalars.
In the same way, one can think of the new first factor as the potential
factor, since to leading order in the low energy expansion, it reproduces the
nonabelian scalar potential expected for the super-Yang-Mills theory --- see
below. Further note that the first factor reduces
to simply one when the scalar fields are commuting, even
for general background fields.

As mentioned below eq.~\reef{csact}, an individual
D$p$-brane couples not only to the RR potential with
form degree $n=p+1$, but also to the RR potentials with $n=p-1,p-3,\ldots$ through
the exponentials of $B$ and $F$ appearing in the Wess-Zumino action \reef{csact}.
Above in eq.~\reef{finalcs}, $\hi_\Phi\hi_\Phi$ is an operator of form degree
--2, and so world-volume interactions appear in the nonabelian action \reef{finalcs}
involving the higher RR forms.
Hence in the nonabelian theory, a D$p$-brane can also couple
to the RR potentials with $n=p+3,p+5,\ldots$ through the additional commutator
interactions. To make these
couplings more explicit, consider the D0-brane action (for which $F$
vanishes):
\bear
S_{CS}&=&\mu_0\int \STr\left(P\left[C^{(1)}+
i\l\,\hi_\Phi \hi_\Phi\left(C^{(3)}+C^{(1)}B\right)
\vphantom{\l^4\over24}\right.\right.
\nonumber\\
&&\quad \qquad-{\l^2\over2}(\hi_\Phi \hi_\Phi)^2\left(C^{(5)}
+C^{(3)}B+{1\over2}C^{(1)}B^2\right)
\nonumber\\
&&\quad -i{\l^3\over6}(\hi_\Phi \hi_\Phi)^3\left(
C^{(7)}+C^{(5)}B
+{1\over2}C^{(3)}B^2+{1\over6}C^{(1)}B^3\right)
\labell{cszero}\\
&&\quad \left.\left.+{\l^4\over24}(\hi_\Phi \hi_\Phi)^4\left(
C^{(9)}+C^{(7)}B+{1\over2}C^{(5)}B^2
+{1\over6}C^{(3)}B^3+{1\over24}C^{(1)}B^4\right)\right]\right).
\nonumber
\eear
Of course, these interactions are
reminiscent of those appearing in Matrix theory \cite{matrix,matbrane}.
For example, eq.~\reef{cszero} includes a linear coupling to $C^{(3)}$, which
is the potential corresponding to D2-brane charge,
\bear
&&i\lambda\,\mu_0\int \Tr\, P\left[\hi_\Phi \hi_\Phi C^{(3)}\right]
\nonumber\\
&&\qquad\quad
=i{\l\over2}\mu_0\int dt\ \Tr \left(C_{tjk}^{(3)}(\Phi,t)\,[\Phi^k,\Phi^j]
+\lambda C^{(3)}_{ijk}(\Phi,t)\,D_t\Phi^k\,[\Phi^k,\Phi^j]
\right)
\labell{magic}
\eear
where we assume that $\sigma^0=t$ in static gauge.
Note that the first term on the right hand side has the form of a source
for D2-brane charge. This is essentially the interaction central
to the construction of D2-branes in Matrix theory with the large N
limit \cite{matrix,matbrane}. Here, however, with finite N,
this term would vanish  upon taking the trace  if
$C_{tjk}^{(3)}$ was simply  a function of the
world-volume coordinate $t$ (since $[\Phi^k,\Phi^j]
\in{\rm SU(N)}\,)$. However, in general
these three-form components  are functionals of $\Phi^i$.
Hence, while there would be no ``monopole'' coupling to D2-brane charge,
nontrivial expectation values of the scalars can give rise to couplings
to an infinite series of higher ``multipole'' moments.

Finally we add that by the direct examination of string scattering amplitudes
using the methods of refs.~\cite{garousi1} and \cite{igor}, one can
verify at low orders the form of the nonabelian interactions
in eqs.~\reef{finalbi} and \reef{finalcs}, including
the appearance of the new commutator interactions in the nonabelian
Wess-Zumino action \cite{garousi3}.

\vskip 3ex
\centerline{\bf 2. Dielectric Branes}   

In this section, we wish to consider certain physical
effects arising from the new nonabelian interactions
in the world-volume action, given by eqs.~\reef{finalbi} and \reef{finalcs}.
To begin,
consider the scalar potential for D$p$-branes in flat space, \ie
$G_{\mu\nu}=\eta_{\mu\nu}$ with all other fields vanishing.
In this case, the entire scalar potential originates in the Born-Infeld
term \reef{finalbi} as
\begin{equation}
V=T_p\,\Tr\sqrt{det(Q^i{}_j)}= \NN T_p-{T_p\l^2\over4}
\Tr([\Phi^i,\Phi^j]\,[\Phi^i,\Phi^j])+\ldots
\labell{trivpot}
\end{equation}
The commutator-squared term
corresponds to the potential for ten-dimensional U(N)
super-Yang-Mills theory reduced to $p+1$ dimensions.
A nontrivial set of extrema of this potential is given by taking the $9-p$ scalars as
constant commuting matrices, \ie
\begin{equation}
\labell{commat}
[\Phi^i,\Phi^j]=0
\end{equation}
for all $i$ and $j$. Since they are commuting, the $\Phi^i$ may
be simultaneously diagonalized and as discussed above,
the eigenvalues are interpreted as the separated positions of N fundamental D$p$-branes
in the transverse space. This solution reflects the fact that a system of
N parallel D$p$-branes is supersymmetric, and so they can sit in
static equilibrium with arbitrary separations in the transverse space
\cite{dbrane1,dbrane2}.

From the results described in the previous section,
it is clear that in going from flat space to general background
fields, the scalar potential is modified by new interactions and so one
should reconsider the analysis of the extrema.
It turns out that this yields an interesting physical effect
that is a precise analog for D-branes of the dielectric effect in ordinary electromagnetism.
That is when D$p$-branes are placed  in a nontrivial background
field for which the D$p$-branes would normally be regarded as neutral,
\eg nontrivial $F^{(n)}$ with $n>p+2$, new terms will be induced
in the scalar potential, and generically one should expect that there
will be new extrema beyond those found in flat space, \ie eq.~\reef{commat}.
In particular, there can be nontrivial extrema
with noncommuting expectation values of the $\Phi^i$, \eg with
$\Tr\Phi^i=0$ but $\Tr(\Phi^i)^2\ne0$. This would correspond to
the external fields ``polarizing'' the D$p$-branes to expand
into a (higher dimensional)
noncommutative world-volume geometry. This is the analog of the familiar
electromagnetic process where an external field may induce
a separation of charges in neutral materials. In this case, the
polarized material will then carry an electric dipole (and possibly
higher multipoles). The latter is also seen in the D-brane analog. When
the world-volume theory is at a noncommutative extremum, the
gauge traces of products of scalars will be nonvanishing in various interactions
involving the supergravity fields. Hence at such an extremum, the
D$p$-branes act as sources for the latter bulk fields.

To make these ideas explicit, we will now illustrate the
process with a simple example. We consider N D0-branes
in a constant background RR field $F^{(4)}$, \ie the field strength
associated with D2-brane charge. We find that the D0-branes expand into
a noncommutative two-sphere which represents a spherical bound state
of a D2-brane and N D0-branes.

Consider a background where only
RR four-form field strength is nonvanishing with
\begin{equation}
F^{(4)}_{tijk}=-2f \vareps_{ijk}\qquad{\rm for}\ i,j,k\in
\lbrace 1,2,3\rbrace
\labell{backg}
\end{equation}
with $f$ a constant (of dimensions $length^{-1}$).
Since $F^{(4)}=dC^{(3)}$,
we must consider the coupling of the D0-branes
to the RR three-form potential, which is given above in eq.~\reef{magic}.
If one explicitly introduces the nonabelian Taylor expansion \reef{slick},
one finds the leading order interaction may be written as
\begin{equation}
{i\over3}\lambda^2\mu_0\int dt\,\Tr\left(\Phi^i\Phi^j\Phi^k\right)
F^{(4)}_{tijk}(t)\ .
\labell{interact3}
\end{equation}
This final form might have been anticipated
since one should expect that the world-volume potential can only depend on
gauge invariant expressions of the background field. Given that
we are considering a constant background $F^{(4)}$,
the higher order terms implicit in eq.~\reef{magic} will
vanish as they can only involve spacetime derivatives of the four-form
field strength. Combining eq.~\reef{interact3} with
the leading order Born-Infeld potential \reef{trivpot} yields the
scalar potential of interest for the present problem
\begin{equation}
V(\Phi)=\NN T_0-{\lambda^2T_0\over4}\Tr([\Phi^i,\Phi^j]^2)
-{i\over3}\lambda^2\mu_0\Tr\left(\Phi^i\Phi^j\Phi^k\right)
F^{(4)}_{tijk}(t)\ .
\labell{potential}
\end{equation}

Substituting in the background field \reef{backg} and $\mu_0=T_0$,
$\delta V(\Phi)/\delta\Phi^i=0$ yields
\begin{equation}
0=[[\Phi^i,\Phi^j],\Phi^j]+{i}\,f\vareps_{ijk}[\Phi^j,\Phi^k]\ .
\labell{eqmot}
\end{equation}
Note that commuting matrices \reef{commat} describing separated D0-branes
still solve this equation.
The value of the potential for these solutions is simply $V_0=\NN T_0$,
the mass of N D0-branes. Another interesting solution of eq.~\reef{eqmot} is
\begin{equation}
\Phi^i={f\over2}\,\alpha^i
\labell{solu1}
\end{equation}
where $\alpha^i$ are any N$\times$N matrix representation of
the SU(2) algebra
\begin{equation}
[\al^i,\al^j]=2i\,\vareps_{ijk}\,\al^k\ .
\labell{su2}
\end{equation}
For the moment, let us focus on the irreducible representation for
which one finds
\begin{equation}
\Tr[(\al_{\ssc \NN}^i)^2]={\NN\over3}(\NN^2-1) \quad{\rm for}\ i=1,2,3.
\labell{trace}
\end{equation}
Now evaluating the value of the potential \reef{potential}
for this new solution yields
\begin{equation}
V_\NN=\NN T_0-{T_0\l^2f^2\over6}\sum_{i=1}^3\Tr[(\Phi^i)^2]
=\NN T_0-{\pi^2\ls^3f^4\over6g}\NN^3\left(1-{1\over\NN^2}\right)
\labell{evpot}
\end{equation}
using $T_0=1/(g\ls)$.
Hence the noncommutative solution \reef{solu1} has lower energy
than a solution of commuting matrices, and so the latter
configuration of separated D0-branes is unstable towards
condensing out into this noncommutative solution.
One can also consider reducible representations of the
SU(2) algebra \reef{su2}, however, one finds that the corresponding energy
is always larger than that in eq.~\reef{evpot}. Hence it seems that
the irreducible representation describes the ground state of the system.

Geometrically, one can recognize the SU(2) algebra
as that corresponding to the noncommutative or fuzzy two-sphere
\cite{hoppen,fuzz}. The physical size of
the fuzzy two-sphere is given by
\begin{equation}
R=\l\left(\sum_{i=1}^3\Tr[(\Phi^i)^2]/\NN\right)^{1/2}=
\pi\ls^2f\NN\left(1-{1\over \NN^2}\right)^{1/2}
\labell{radius}
\end{equation}
in the ground state solution.
From the Matrix theory construction of Kabat
and Taylor \cite{wati2}, one can infer this ground state is not simply
a spherical arrangement of D0-branes rather the noncommutative solution actually
represents a spherical D2-brane with N D0-branes bound to it.
In the present context, the latter can be verified by seeing that this
configuration has a ``dipole'' coupling to the RR four-form.
The precise form of this
coupling is calculated by substituting the noncommutative
scalar solution \reef{solu1} into the world-volume interaction
\reef{interact3}, which yields
\begin{equation}
-{R^3\over3\pi g\ls^3}\left(1-{1\over \NN^2}\right)^{-1/2}\int dt\,F^{(4)}_{t123}\ .
\labell{dipole}
\end{equation}
for the ground state solution. Physically this $F^{(4)}$-dipole moment arises
because antipodal surface elements on the sphere have the opposite
orientation and so form small pairs of separated
membranes and anti-membranes. Of course, the spherical configuration
carries no net D2-brane charge.

Given that the noncommutative ground state solution corresponds to a bound
state of a spherical D2-brane and N D0-branes, one might attempt to match
the above results using the dual formulation. That is, this system can
be analyzed
from the point of view of the (abelian) world-volume theory of a D2-brane.
In this case, one would consider a spherical D2-brane carrying a
flux of the U(1) gauge field strength representing the N bound D0-branes,
and at the same time, sitting in the background of the constant RR four-form
field strength \reef{backg}.
In fact, one does find stable static solutions, but what is more
surprising is how well the results match those calculated in the framework
of the D0-branes. The results for the energy, radius and
dipole coupling are the same as in eqs.~\reef{evpot}, \reef{radius}
and \reef{dipole}, respectively, except that the factors of $(1-1/\NN^2)$
are absent \cite{die}. Hence for large N, the two calculations agree up to
$1/\NN^2$ corrections.

One expects that the D2-brane calculations would be valid when $R\gg\ls$
while naively the D0-brane calculations would be valid when $R\ll\ls$.
Hence it appears there is no common domain where the two pictures can
both produce reliable results. However, a more careful consideration
of range of validity of the D0-brane calculations only requires that
$R\ll\sqrt{\NN}\ls$. This estimate is found by requiring that the
scalar field commutators appearing in the full nonabelian potential
\reef{trivpot} are small so that the Taylor expansion of the square root
converges rapidly. Hence for large N, there is a large domain of
overlap where both of the dual pictures are reliable.
Note the density of D0-branes on the two-sphere is
$\NN/(4\pi R^2)$. However, even if $R$ is macroscopic it is still bounded by
$R\ll\sqrt{\NN}\ls$ and so this density must
be large compared to the string scale, \ie the density is much larger than
$1/\ls^2$. With such large densities, one can imagine the discreteness of the fuzzy
sphere is essentially lost and so there is good agreement with the continuum
sphere of the D2-brane picture.

Finally note that the Born-Infeld action  contains couplings to the
Neveu-Schwarz two-form which are similar to that in eq.~\reef{interact3}.
{}From the expansion of $\sqrt{det(Q)}$, one finds a cubic interaction
\begin{equation}
{i\over3}\lambda^2T_0\int dt\,\Tr\left(\Phi^i\Phi^j\Phi^k\right)
H_{ijk}(t)\ .
\labell{nese3}
\end{equation}
Hence the noncommutative ground state,
which has $\Tr\left(\Phi^i\Phi^j\Phi^k\right)\ne0$, also
acts as a source of the $B$ field with
\begin{equation}
-{R_0^3\over3\pi g\ls^3}\left(1-{1\over \NN^2}\right)\int dt\,H_{123}\ .
\labell{hdipole}
\end{equation}
This coupling is perhaps not so surprising given that the noncommutative
ground state represents the bound state of a spherical D2-brane
and N D0-branes. Explicit supergravity solutions describing D2-D0
bound states with a planar geometry have been found \cite{useful,russo},
and are known to carry a long-range $H$ field with the same profile as the
RR field strength $F^{(4)}$. One can also derive this coupling
from the dual D2-brane formulation.
Furthermore, we observe that the presence of this coupling \reef{nese3}
means that we would find an analogous dielectric effect if
the N D0-branes were placed in a constant background $H$ field.

The example considered above must be considered simply a toy calculation
demonstrating the essential features of the dielectric effect for D-branes.
A more complete calculation would require analyzing the D0-branes in a
consistent supergravity background. For example, the present case could be
extended to consider the asymptotic supergravity fields of a D2-brane, where
the RR four-form would be slowly varying but the metric and dilaton
fields would also be nontrivial. Alternatively, one can find solutions with
a constant background $F^{(4)}$ in M-theory, namely the AdS$_4\times$S$^7$
and AdS$_7\times$S$^4$ backgrounds --- see, \eg \cite{duff}. In lifting the
D0-branes to M-theory, they become gravitons carrying momentum in the internal space.
Hence the expanded D2-D0 system considered here correspond to the ``giant
gravitons'' of ref.~\cite{giant}. The analog of the D2-D0 bound state in
a constant background $F^{(4)}$ corresponds to M2-branes with internal
momentum expanding into AdS$_4$ \cite{goliath,aksun}, while the that in
a constant $H$ field corresponds to the M2-branes expanding on S$^4$ \cite{giant}.
Alternatively, the dielectric effect has been found to play a role in other
string theory contexts, for example, in the resolution of certain
singularities in the AdS/CFT correspondence \cite{joemat}, or in
describing D-branes in the spacetime background corresponding to a WZW model
\cite{WZW1,WZW2}. Further, one can consider more sophisticated background
field configurations which through the dielectric effect generate more
complicated noncommutative geometries \cite{sand}.

\vskip 3ex
\centerline{\bf 3. Giant Gravitons} 

From the above discussion, it seems that in the M-theory
backgrounds of AdS$_{4}\times \S^7$ or AdS$_{7}\times \S^4$,
one will find that an M2-brane carrying internal momentum will
expand into a stable spherical configuration. While a Matrix
theory description of such states in terms of noncommutative geometry
is not yet possible, one can instead analyse these configurations
in terms of the abelian world-volume theory of the M2-brane.
In fact, the spherical
M2-branes expanding into AdS$_4$ were actually discovered
some time ago \cite{oldm}. It turns out
that M5-branes will expand in a similar way for these backgrounds,
and further that expanded D3-branes arise in the type IIB supergravity
background AdS$_{5}\times \S^5$. A detailed analysis
\cite{goliath,giant,aksun}
shows that these expanded branes are BPS states with the quantum numbers
of a graviton. In the following, we will discuss
the details of the effect for the D3-branes.
Most of the discussion applies equally well for the analogous
M2- and M5-brane configurations.

The line element for $\AdS_5\times \S^5$ may be written as:
\beq
ds^2=-\left(1+{r^2\over L^2}\right)dt^2+{dr^2\over
1+{r^2\over L^2}}+r^2 d\Omega_{3}^2+
L^2\left(d\theta^2+\cos^2\theta d\phi^2+\sin^2\theta
d\tOmega_{3}^2\right)\ .
\labell{metric}
\eeq
This background also involves a self-dual RR five-form field strength 
with terms proportional to the volume forms on the two five-dimensional 
subspaces: $F^{(5)}= \frac{4}{L}[\varepsilon({\rm AdS}_5)+\varepsilon( \S^5)]$.
With the coordinates chosen above, 
the four-form potential on the the AdS part of the space is
\beq
C^{(4)}_{electric}=-{r^{4}\over
L}dt\,\vareps(\S^{3})
\labell{adsp}
\eeq
where $\vareps(\S^{3})$ is the volume form for the
three-sphere described by $d\Omega_{3}^2$. Similarly, the potential
on the $\S^5$ is
\beq
C^{(4)}_{magnetic} = L^{4} \sin^{4}\theta\,d\phi\,\vareps(\tilde{\S}^3)
\labell{sphp}
\eeq
where $\vareps(\tilde{\S}^3)$ is the volume form on $d\tOmega_{3}^2$.
For the D3-brane configurations of interest, the world-volume action
in eqs.~\reef{biact} and \reef{csact} reduces to:
\beq
S_{3}=-T_3\int d^{4}\sigma\ \sqrt{-det(P[G])}+T_3\int P[C^{(4)}]\ .
\labell{actp}
\eeq
Here, the world-volume gauge field has been set to zero, which will
be consistent with the full equations of motion.

Following ref.~\cite{giant}, one can find solutions where a D3-brane
has expanded on the $\S^5$ to a sphere of fixed $\theta$ while it orbits
the $\S^5$ in the $\phi$ direction. Our static gauge choice matches
the spatial world-volume coordinates with the angular coordinates on
$d\tOmega^2_3$, and identifies $\s^0=t$.
Now we consider a trial solution of the form:
$\theta = {\rm constant}$, $r=0$ and $\phi=\phi(t).$
Substituting this ansatz into the world-volume action 
\reef{actp} and integrating over the angular coordinates, yields the
following Lagrangian
\beq
\cL_{3}=\frac{\NN}{ L}\left[-\sin^{3}\theta\,\sqrt{1-L^2 \cos^2 \theta 
\,\phd^2}+
L\sin^{4} \theta \,\phd\right]\ .
\labell{lag2}
\eeq
Here we have introduced the (large positive) integer $\NN$ which counts
the five-form flux on $\S^5$. This is also, of course, the rank of
the U(N) gauge group in the dual super-Yang-Mills theory.
Introducing the conjugate angular momentum
$P_\phi=\delta\cL_{3}/\delta\phd$, we construct the Hamiltonian:
\beq
\cH_{3}=P_\phi\phd-\cL_{3}
={\NN\over L}\sqrt{p^2+\tan^2\theta\,(p-\sin^{2}\theta)^2}
\labell{ham2}
\eeq
where $p=P_\phi/\NN$.
Given that the Hamiltonian is independent of $\phi$,
the equations of motion will be solved with constant angular momentum (and
hence constant $\phd$). For
fixed $p$, eq.~\reef{ham2} can be regarded as the  potential that determines
the angle $\theta$ for equilibrium. Examining $\cH_{3}$ in detail reveals
degenerate minima at $\sin\theta=0$ and $\sin^2\theta=p$, and
at any of these minima, the energy is $\cH_{3}=\Pp/L$.
The expanded configurations are then the giant gravitons
of ref.~\cite{giant}. An important observation is that
the minima at $\sin^2\theta=p$ only exist
for $p\le1$. As $p$ grows beyond $p=1$, the minima at $\theta\ne0$
are lifted above that at $\sin\theta=0$ and then disappear
completely if $p>9/8$.

The discussion above indicates that one can also consider
the possibility of a brane expanding into the AdS part of the spacetime
\cite{goliath,aksun}.
That is we wish to find solutions where a D3-brane has 
expanded to a sphere of constant $r$ while it still orbits in the 
$\phi$ direction on the $\S^5$. Choosing static gauge, we  again identify
$\sigma^0= t$ but match the remaining world-volume coordinates
with the angular coordinates on $d\Omega_{3}^2$.
The trial solution is now: $\theta = 0$, $r={\rm constant}$ and
$\phi=\phi(t)$. Beginning with the same
world-volume action \reef{actp} \cite{foot},
one calculates as before and arrives at the following Hamiltonian
\beq
\cH_{3}
={\NN\over L}\left[
\sqrt{\left(1+{r^2\over L^2}\right)\left(p^2+{r^{6}\over L^6}
\right)}-{r^{4}\over  L^4}\right]\ .
\labell{green2}
\eeq
where as before $p=\Pp/\NN$.
Examining $\prt \cH_3/\prt r=0$, one finds minima
located at $r=0$ and $\left(r/L\right)^2=p$.
The energy at each of the minima is $\cH_{3}=P_\phi/L$.
In ref.~\cite{goliath}, these expanded configurations were
denoted as dual giant gravitons.
An essential difference from the previous case, however, is that
the minima corresponding to expanded branes persist for arbitrarily large 
$p$.

It is interesting to consider the motion of these expanded brane
configurations.
Evaluating $\phd$ for any of the above solutions,
remarkably one finds the same result: $\phd=1/L$, independent of $\Pp$.
Further the center of mass motion for any of the equilibrium
configurations in the full 
ten-dimensional background is along a null trajectory. For example,
for the D3-branes expanded on $\S^5$
\beq
ds^2=-(1-L^2\,\cos^2\theta\,\phd^2)dt^2=0
\labell{null}
\eeq
when evaluated for $\phd=1/L$ and $\theta=0$(= the center of mass position).
This is, of course, the expected result for a massless `point-like'
graviton, but it applies equally well for both
of the expanded brane configurations.
However, note that in the expanded configurations, the motion
of each element of the sphere is along a time-like trajectory.

From the point of view of five-dimensional supergravity in the
AdS space, the stable brane configurations correspond to massive states with
$M = P_\phi/L$. Their angular momentum means that these states are
also charged under a U(1) subgroup of the SO(6) gauge symmetry in
the reduced supergravity theory. With the appropriate normalizations,
the charge is $Q= P_\phi/L$, and hence one finds that these
configurations satisfy the appropriate BPS bound \cite{giant}.
One can therefore anticipate that all of these configurations should
be supersymmetric. The latter result has been verified by an explicit
analysis of the residual supersymmetries \cite{goliath,aksun}.

The AdS$_5 \times \S^5$ background is a maximally 
supersymmetric solution of the type IIB supergravity equations with 32
residual supersymmetries. That is the background fields are invariant under
supersymmetries parameterized by 32 independent Killing spinors. These
Killing spinors are determined by setting
\beq
\delta \Psi_M = {D}_M \eps - \frac{i}{480}\,{\Gamma_M}^{PQRST}
F^{(5)}_{PQRST}\,\eps=0
\eeq
as the variations of all of the other type IIB supergravity fields vanish
automatically. The solutions take the form $\eps=M(x^\mu)\eps_0$ 
where $\eps_0$ is an arbitrary constant complex Weyl spinor.

A supersymmetric extension of the abelian world-volume action
has been constructed for D3-branes
(and all other D$p$-branes) in a general supergravity background
\cite{3brane,3braneb}.
This action can be viewed as a four-dimensional nonlinear sigma
model with a curved superspace as the target space. Hence the theory
is naturally
invariant under the target-space supersymmetry. Further however,
formulating the action with manifest ten-dimensional Lorentz invariance,
requires an additional fermionic invariance on the world-volume called
$\kappa$-symmetry. For a test brane configuration where both the
target space and world-volume fermions vanish, residual supersymmetries
may arise provided there are Killing spinors which satisfy a combined
target-space supersymmetry and $\kappa$-symmetry transformation. The
latter amounts to imposing a constraint $\Gamma\eps=\eps$ where
\beq
\Gamma=-\frac{i}{4!}\veps^{i_1\cdots i_4} \prt_{i_1}X^{M_1}
\cdots\prt_{i_4}X^{M_4}\Gamma_{M_1\cdots M_4}\ .
\labell{lousya}
\eeq
Of course, this constraint is only evaluated on the D3-brane
world-volume. For all of the minima of the potentials in both
eqs.~\reef{ham2} or \reef{green2},
this constraint reduces to imposing the same projection
\beq
(\Gamma^{t\phi}+1)\eps_0=0\ .
\labell{endresult}
\eeq
Hence not only are the expanded branes and the point-like state all BPS 
configurations, all of these configurations preserve precisely the same
supersymmetries. Note that this projection is what one might
have expected for a massless particle moving along the $\phi$ direction,
\eg one can compare to the supersymmetries gravitational waves propagating in
flat space \cite{wave}.

Much of the interest in  giant gravitons comes
from an intriguing suggestion \cite{giant} that they may be related to the
`stringy exclusion principle' \cite{exclus,exclusb,exclusc,exclusd}.
The latter arises
in the AdS/CFT correspondence \cite{revue}\ where it is easily
understood in the conformal field theory. A family
of chiral primary operators in the $N$=4 super-Yang-Mills theory
terminates at some maximum weight because the U(N) gauge group has a
finite rank. In terms of the dual AdS description, these operators are
associated with single particle states carrying angular
momentum on the internal five-sphere. So the appearance of an upper 
bound on the angular momentum seems mysterious from the point of view of the 
supergravity theory. The suggestion of McGreevy, Susskind and Toumbas
\cite{giant} is that if the dual single particle states are identified
with the giant gravitons, the D3-branes expanded on the $\S^5$,
then the upper bound is
produced by the fact that these BPS states only exist for $p\le1$.
In fact, this exactly reproduces the desired upper bound
on the angular momentum: $P_\phi\le\NN$.

Unfortunately this interpretation is not entirely clear because
rather than a unique candidate for the graviton state, there
are {\it three} different ones, including the giant gravitons
which expand on $\S^5$, the dual giant gravitons which expand
on AdS$_5$, and the point-like states. All of these configurations
have the same angular momentum and energy, and preserve precisely
the same supersymmetries. Unfortunately the latter two of the candidates
display no upper bound on the angular momentum, and so there is some
uncertainty about the proposed mechanism for the stringy
exclusion principle. 

One tentative suggestion \cite{goliath} is that the exclusion principle
may be realized through the quantum mechanical mixing of these different
states. One can find instanton configurations describing tunneling
between the point-like states and either of the expanded branes
\cite{goliath,aksun}, but not between the two expanded D3-brane solutions
\cite{tunnel}. The suggestion is then that this mixing may
spontaneously break supersymmetry in the regime $P_\phi>\NN$ where there are
only two potential graviton states.

Ref.~\cite{aksun} has done some interesting calculations in the
context of the dual CFT. They seem to be able to identify
certain classical field configurations with same properties as
the dual giant gravitons. Further these calculations seem
to indicate that the minimum corresponding to the point-like
graviton is lifted due to strong coupling effects. This then
suggests a picture where the D3-branes expanded on AdS$_5$
are dual to coherent states in the $N$=4 super-Yang-Mills theory,
and so do not directly correspond to the chiral primary operators
considered in the stringy exclusion principle.

\vskip 3ex
\centerline{\bf 4. Intersecting Branes} 

One interesting aspect of the (abelian) Born-Infeld action \reef{biact}
is that it supports solitonic configurations describing
lower-dimensional branes protruding from the original D-brane
\cite{calmald,gibb,selfd}. For example, in the case of a D3-brane,
one finds spike solutions, known as ``bions,'' corresponding to
fundamental strings and/or D-strings extending out of the D3-brane.
In these configurations, both the
world-volume gauge fields
and transverse scalar fields are excited. The gauge field corresponds to
that of a point charge arising from the end-point of the attached
string, {\it i.e.}, an electric charge for a fundamental
string and a magnetic monopole charge
for a  D-string. The scalar field describes
the deformation of the D3-brane geometry caused by attaching the strings.
These solutions seem to have a surprisingly wide range of validity, even
near the core of the spike where the fields are no longer slowly varying.
In fact, one can show that the electric spike corresponding to a fundamental
string is a solution of the full string theory equations of motion \cite{lars}.
Further the dynamics of these solutions, as probed through small fluctuations,
agrees with the expected string behavior \cite{fluc1,fluc2,fluc3,fluc4}.
In part, these remarkable agreements are probably related to the fact that
these are supersymmetric configurations.

For the system of N D-strings ending on a D3-brane, there is also
a dual description in terms of the nonabelian world-volume theory of
the N D-strings. There one finds solutions which have an interpretation,
in terms of noncommutative geometry, as describing the D-strings expanding
out in a funnel to become an orthogonal D3-brane.
In fact, there is an extensive discussion of this system in the literature
--- see, \eg
\cite{ded,gauntlett,brech,giveon,sethi,tsim,hashimoto,gorsky} --- where the
emphasis
was on the close connection \cite{ded} of the D-string equations to
the Nahm equations for BPS monopoles \cite{nahm}.
In ref.~\cite{bion}, our emphasis was on the interpretation of these
solutions in terms of noncommutative geometry and the remarkable agreement
that one finds with the D3-brane spikes in the large N limit.

For N D-strings in flat space, the dynamics is determined completely by the
Born-Infeld action (\ref{finalbi}) which reduces to \cite{die,ark2}
\begin{equation}
S=-T_1\int d^2\sigma\, \STr\sqrt{-\det\left(\eta_{ab}+
\lambda^2\partial_a\Phi^i Q^{-1}_{ij}\partial_b\Phi^j\right)
\ \det\left(Q^{ij}\right)}\ ,\labell{action2}
\end{equation}
where
\begin{equation}
Q^{ij}=\delta^{ij}+i\lambda[\Phi^i,\Phi^j]\ .
\end{equation}
Implicitly here, the world-volume gauge field has been set to zero,
which will be a consistent truncation for the configurations considered below.
With the usual choice of static gauge, we set
the world-volume coordinates: $\tau=t=x^0$ and $\sigma=x^9$.
For simplicity, one might consider the leading-order (in $\lambda$)
equations of motion coming from this action:
\begin{equation}
\prt^a\prt_a\Phi^i=[\Phi^j,[\Phi^j,\Phi^i]]\ .
\labell{motion}
\end{equation}
Of course, a simple solution of these equations are constant commuting matrices,
as in eq.~\reef{commat}. As discussed in the previous section, such
a solution describes N separated parallel D-strings sitting in static
equilibrium.

To find a dual description of the bion solutions of the D3-brane
theory \cite{calmald,gibb}, one needs a static solution which represents the D-strings
expanding into a D3-brane. The corresponding geometry would
be a long funnel where the cross-section at fixed $\sigma$
has the topology of a two-sphere. In this context, the latter cross-section
naturally arises as a fuzzy two-sphere \cite{hoppen,fuzz} if the scalars have
values in an N$\times$N matrix representation of the SU(2) algebra \reef{su2}.
Hence one is lead to consider the ansatz
\begin{equation}
\Phi^i = {R(\sigma)\over\l\sqrt{\NN^2-1}}\,\alpha^i,\ \ i=1,2,3,\labell{ansatz}
\end{equation}
where we will focus on case where the $\alpha^i$ are the irreducible $N\times N$
SU(2) matrices. Then with the normalization in eq.~\reef{ansatz}, the function
$|R(\sigma)|$ corresponds precisely to the radius of the fuzzy two-sphere
\begin{equation}
R(\sigma)^2={\l^2\over \NN}\sum_{i=1}^3\Tr[\Phi^i(\sigma)^2]\ .
\labell{radii}
\end{equation}
Substituting the ansatz \reef{ansatz} into the matrix equations of motion
\reef{motion} yields a single scalar equation
\begin{equation}
R''(\sigma)={8\over\l^2(\NN^2-1)} R(\sigma)^3\ ,\labell{leom}
\end{equation}
for which one simple class of solutions is
\begin{equation}
R(\sigma)=\pm {\NN\pi\ls^2\over\sigma-\signot}\left(1-{1\over \NN^2}\right)^{1/2}\ .
\labell{spike}
\end{equation}

Given the above analysis, eqs.~\reef{ansatz} and \reef{spike} only represent a
solution of the leading order equations of motion \reef{motion},
and so naively one expects that it should only be valid for small
radius. However, one can show by direct evaluation
\cite{bion} that in fact these configurations
solve the full equations of motion extremizing the nonabelian action
\reef{action2}. The latter can also be inferred from an analysis
of the world-volume supersymmetry of these configurations. Killing spinor solutions of
the linearized supersymmetry conditions will exist provided
that the scalars satisfy
\begin{equation}
D_\sigma \Phi^i=\pm {i\over2}\varepsilon^{ijk}\left[\Phi^j,
\Phi^k\right]\ .\label{nahmeq}
\end{equation}
The latter can be recognized as the Nahm equations \cite{ded}. Hence
the duality between the D3-brane and D-string descriptions gives a physical
realization of Nahm's transform of the moduli space of BPS magnetic monopoles.
Now inserting the ansatz (\ref{ansatz}) into eq.~\reef{nahmeq} yields
\begin{equation}
R'=\mp {2\over\l\sqrt{N^2-1}}R^2\labell{susyeq}
\end{equation}
which one easily verifies is satisfied by the configuration
given in eq.~\reef{spike}. Hence, one concludes
that the solutions given by eqs.~\reef{ansatz} and
\reef{spike} are in fact BPS solutions preserving $1/2$ of the supersymmetry
of the leading order D-string theory. Now in ref.~\cite{aki}, it was shown
that BPS solutions of the leading order theory
are also BPS solutions of the full nonabelian Born-Infeld action
\reef{action2}.

The geometry of the solution, eqs.~\reef{ansatz} and
\reef{spike}, certainly has the desired funnel shape.
The fuzzy two-sphere shrinks to zero size as $\sig\rightarrow\infty$
and opens up to fill the 
$x^{1,2,3}$ hypersurface
at $\sig=\signot$. By examining the nonabelian Wess-Zumino action \reef{finalcs},
one can show that the noncommutative solution induces a coupling to the RR
four-form potential $C^{(4)}_{t123}$. This calculation confirms then that, with
the minus (plus) sign in eq.~\reef{spike},
the D-strings expand into a(n anti-)D3-brane which fills the
$x^{1,2,3}$ directions \cite{bion}. Given that the funnel solution of the
D-string theory and the bion spike of the D3-brane theory are both
BPS, one might expect that there will be a good agreement between these
two dual descriptions. The
formula for the height of D3-brane spike above the $x^{1,2,3}$ hyperplane
is \cite{calmald}
\begin{equation}
\sigma-\signot={\NN\pi \ls^2\over R}\ .\labell{reverse}
\end{equation}
Comparing to eq.~\reef{spike}, one finds that for large N the two descriptions
are describing the same geometry up to $1/\NN^2$ corrections. One finds similar
quantitative agreement for large N in calculating the energy, the RR couplings
and the low energy dynamics in the two dual descriptions \cite{bion}.
As in the discussion
of the dielectric effect, one can argue that the D3-brane description is valid
for $R\gg\ls$ while the D-string description is reliable for $R\ll\sqrt{\NN}\ls$
\cite{bion}. Hence one can understand the good agreement between these dual
approaches for large N since there is a large domain of overlap where both are
reliable.

Note that in the configurations considered in this section,
there are no nontrivial supergravity fields in the ambient spacetime.
Hence the appearance of the noncommutative geometry in these solutions
is quite distinct from that in the dielectric effect, where the external
fields drive the D-branes into a certain geometry in the ground state.
In the funnel solutions, the noncommutative geometry was put into the ansatz
\reef{ansatz} by hand. An interesting extension of these solutions is then
to replace the SU(2) generators by those corresponding to some other
noncommutative geometry, \ie to replace eq.~\reef{ansatz} by
\begin{equation}
\Phi^i={R(\sig)\over\l\sqrt{C}}\,G^i
\labell{newansz}
\end{equation}
where the $G^i$ are new N$\times$N constant matrices satisfying $\sum (G^i)^2 = N\,C$.
An interesting feature of such a construction is that near the
core of the funnel, the leading order equations of motion will still be
those given in eq.~\reef{motion}. Thus for eq.~\reef{newansz} to provide a solution,
the new generators must satisfy $[G^j,[G^j,G^i]]=2a^2\,G^i$ for some constant $a$,
and then the radius is determined by
\begin{equation}
R''= {2a^2\over\l^2C} R^3\ ,
\labell{newdiffer}
\end{equation}
which still has essentially the same form as eq.~\reef{leom} above.
Further the funnel solution of this equation
also has essentially the same form as eq.~\reef{spike}
above, \ie
\begin{equation}
R=\pm{\l\sqrt{C}\over a(\sig-\signot)}\ .
\labell{newsolll}
\end{equation}
Hence the profile with $R\simeq\l/\sig$ is
universal for all funnels on the D-string, independent of the details of the
noncommutative geometry that describes the cross-section of the funnel.

This universal behavior is curious.
For example, one could consider using this framework to describe a D-string ending
on an orthogonal D$p$-brane with $p>3$. However, from the dual D$p$-brane formulation,
one expects that for large $R$, solutions will
essentially be harmonic functions
behaving like $\sig\propto R^{-(p-2)}$ or $R\propto\sig^{-1/(p-2)}$.
The resolution of this puzzle seems to be that
the two profiles apply in distinct regimes, the first for
small $R$ and the second for large $R$. Hence it must be that the nonlinearity
of the full Born-Infeld action will generate solutions which
display a transition from one kind of behavior to another.

One particular example that we have examined in detail \cite{fiv} is the case
where $G^i$ in eq.~\reef{newansz} are chosen to be generators describing a
fuzzy four-sphere --- these may be found in, \eg ref.~\cite{wati2}. In this
case, the funnel describes the D-strings expanding into a D5-brane. One does
find the expected transition in the behavior of the geometry. That is,
$\sig\approx\NN^{2/3}\ls/R$ for small $R$ in accord with eq.~\reef{newsolll},
while at large $R$, higher order terms in the Born-Infeld action \reef{action2}
become important yielding $\sig\approx\NN^{2/3}\ls^4/R^3$. The same kind of
behavior is also found for the corresponding solutions in the dual D5-brane
world-volume theory, although
of course in that case the nonlinearities of the Born-Infeld action become
important for small $R$.
An interesting feature of the D5-brane spike is that
it is also nonabelian in character. Charge conservation arguments
indicate that the D-string acts as a source of the second Chern
class in the world-volume of the D5-brane \cite{semenoff}.
More precisely,
if N D-strings end on a D5-brane, then
\begin{equation}
\label{chargeeq}
{1\over 8\pi^2}\int_{S^4}\Tr(F\wedge F)=\NN\ ,
\end{equation}
for any four-sphere surrounding the D-string endpoint. Hence
both of the dual descriptions have a noncommutative character.
Again, we find that the dual constructions seem to agree at large N,
however, the details of the solutions are more complex. In part, the latter
must be due to the fact that the $D5\perp D1$ system is not supersymmetric.

\vskip 3ex
\centerline{\bf Acknowledgments}
This research was supported by NSERC of Canada and Fonds FCAR du Qu\'ebec.
I would like to thank Neil Constable, Marc Grisaru and \O yvind Tafjord for
collaborations in the research presented in refs.~\cite{goliath}, \cite{bion}
and \cite{fiv}. I would also like to thank Neil Constable and 
\O yvind Tafjord for proofreading a draft of this paper. I would like to
thank the organizers of the Workshop on
Strings, Duality and Geometry (at Universit\'e de Montreal,
March 2000) and of Strings 2000 (at University of Michigan, July  2000)
for the opportunity to speak on the material discussed in this article.
Finally I would like to thank Mike Duff for the invitation to contribute
to this special issue of Journal of Mathematical Physics devoted to Strings,
Branes and M-theory.


\begin{thebibliography}{99}
\bibitem 
{foot}{Our conventions are such that we actually consider an
anti-D3-brane here \cite{goliath}. That is the  sign of the Wess-Zumino
term in eq.~\reef{actp} was reversed.}
\end{thebibliography}
\end{document}